\documentclass[conference]{IEEEtran}

\usepackage{rotating}
\usepackage{cite,xurl}
\usepackage{amsmath,amssymb,amsfonts}
\usepackage{algorithmic}
\usepackage{graphicx}
\usepackage{textcomp}
\usepackage{algorithm2e}
\usepackage{xcolor}
\def\BibTeX{{\rm B\kern-.05em{\sc i\kern-.025em b}\kern-.08em
    T\kern-.1667em\lower.7ex\hbox{E}\kern-.125emX}}

\usepackage{caption}
\usepackage{subcaption}
    
\usepackage{tikz}
\usetikzlibrary{quantikz}

\usepackage{subcaption}

\usepackage{amsmath}
\usepackage{amssymb}
\usepackage{amsthm}

\newtheoremstyle{ieeestyle}
  {\topsep}   % ABOVESPACE
  {\topsep}   % BELOWSPACE
  {\itshape}  % BODYFONT (italicized)
  {}          % INDENT (empty for no indent)
  {\bfseries} % HEADFONT (bold)
  {.}         % HEADPUNCT (punctuation after theorem)
  { }         % HEADSPACE (space after theorem)
  {}          % CUSTOM-HEAD-SPEC

\theoremstyle{ieeestyle}

\begin{document}

\title{Encodings and Node Architecture: Lessons from the Fault-Tolerant Distributed Transversal Fanout}

\author{\IEEEauthorblockN{Seng W. Loke}
\IEEEauthorblockA{\textit{School of Information Technology, Deakin University, Burwood, VIC 3125, Australia.} \\
seng.loke@deakin.edu.au}
 }

%\and

\maketitle

\begin{abstract}
An encoding choice specifies the quantum error-correcting code used to encode logical qubits at a given node. Different encoding choices may differ in the number of physical qubits, the number of encoded logical qubits, the code distance, and the available fault-tolerant operations.
We study different encodings for a distributed transversal fanout operation and look at its implications on node architecture. Our experiments suggest that an encoding with higher distance, even if requiring more ebits, can yield a lower logical error rate for distributed operations, and that  ``upgrading'' an encoding prior to distributed operations and having a  cache, as part of node architecture, for qubits   can be useful for  distributed operations.
\end{abstract}
%why global gates? universl and effiicen ton some computers and a pathological/worse case for DQC!!!


\section{Introduction}
 {\em Fanout} operations, with one control qubit for multiple target qubits, are amenable to efficient implementation in some types of quantum hardware such as trapped-ion quantum computers, enabling reduced depth quantum computations~\cite{fenner23,hoyer05}.  It has also been considered in other platforms, e.g.,~\cite{PhysRevA.105.042404,Liu2026LinearOF}.
 
In the context of distributed or modular quantum computing, often touted as a means of scaling up quantum computation, beyond  monolithic platforms~\cite{BARRAL2025100747,knörzer2025,CALEFFI2024110672,arquin},\footnote{For example, see ~\cite{photonics}, \url{https://spectrum.ieee.org/quantum-computers}, \url{https://newsroom.ibm.com/2025-11-20-ibm-and-cisco-announce-plans-to-build-a-network-of-large-scale,-fault-tolerant-quantum-computers}. See also \url{https://quantnet.lbl.gov} and \url{https://www.ox.ac.uk/news/2025-02-06-first-distributed-quantum-algorithm-brings-quantum-supercomputers-closer}.}
 distributed fanout gadgets using distributed GHZ states have been proposed in~\cite{Yimsiriwattana:2004xhy}, and investigated in~\cite{11662257,11500410}.
Towards fault-tolerant distributed quantum computing, different encodings for quantum error correction  have been proposed and are being implemented,   including surface codes (e.g., ~\cite{singh2025modular,Chandra:2026ovs,ds45-fm9n}), Bivariate-Bicycle (BB) codes (e.g., transversal non-local CNOTs~\cite{Stack2026Transversal} and inter-module operations~\cite{yoder2025tour}), and floquet codes~\cite{sutcliffe2025distributed}. A fault-tolerant distributed 
transversal fanout was  discussed in~\cite{transversalfanout}. Previous work looked at encoding choices for transversal non-local CNOTs and corresponding LER estimates for   distributed global gates based on different encodings~\cite{Loke2026Encoding}.

This paper studies 
transversal implementations of fault-tolerant  distributed fanout with qubits in different   Bivariate-Bicycle (BB) encodings. Our results suggest (i) that with good enough encodings and low enough physical error rates, distributed quantum operations can achieve comparable Logical Error Rates (LERs) with equivalent (entirely) local computations, and (ii) that even if an encoding uses more physical qubits causing non-local transversal operations to consume more ebits, this can be compensated for, in terms of LER, by the higher code distance. We also explore the implications of the results for designing nodes participating in distributed  quantum computations.

In the rest of this paper, we first briefly review  distributed fanout operations in \S2, and then look at distributed or non-local transversal fanout operations in \S3. In \S4, we provide a simulation analysis of distributed transversal fanouts with different encodings, and then explore implications for node architecture design in \S5.   We  conclude with avenues for future work in \S6.

\section{Brief Review of Distributed Fanout Operations}

\begin{figure}
\centering
  \includegraphics[width=0.34\textwidth]{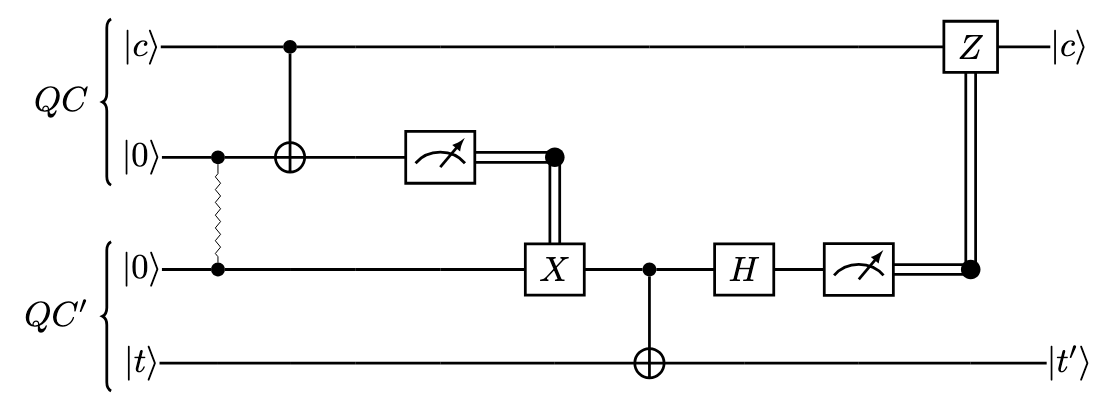}        $\equiv$ 
  \scalebox{0.8}{\begin{quantikz}  c~ & \ctrl{1}\gategroup[2,steps=1,style={dashed,inner sep=6pt}]{Dc-U}   & \qw  \\ 
t  & \gate{U}   & \qw    \end{quantikz} }
  \caption{Distributed controlled-$U$ (i.e., dCNOT is when U=$X$ ($\oplus$) as shown)   between nodes $QC$ and $QC'$ (the {\em computation} qubits are $c$, the control qubit, and $t$, the target qubit), and the wavy line illustrates a Bell pair  involving  the  {\em communication} qubits from the two nodes both initially $\ket{0}$. The right hand side shows the notation we will use for a distributed controlled-$U$ operation.}
\label{dcnot}
\end{figure}

We first consider the distributed CNOT gate (or  dCNOT, in short, a.k.a. non-local CNOT), which is needed as a resource for each dCNOT gate, as shown in Figure~\ref{dcnot}.

\begin{figure*}

\begin{center}
\scalebox{.9}{
      \begin{quantikz} 
   \lstick[2]{A'}    c~                           \qw   & \qw  & \ctrl{1}  & \qw   & \qw & \qw & \qw &   \qw &  \qw   &   \gate{Z^{m_1 \oplus m_2 \oplus m_3}} & \qw c  \\
    a_0    \qw & \ctrl{}{}  & \targ{}  & \qw & \meter{} &  \cwbend{5} &    m_0      &            &  &         \\
    \lstick[2]{A}    a_1    \qw & \ctrl{}{}  & \qw       & \qw & \qw      & \gate{X^{m_0}}   & \ctrl{1} & \gate{H}& \meter{} &  \cwbend{-2} &  m_1 \\
      t1   \qw & \qw        & \qw       & \qw & \qw      & \qw         & \gate{U_1}       & \qw        & \qw &  \qw &\qw   \\
          \lstick[2]{B}    a_2   \qw & \ctrl{}{}  & \qw       & \qw & \qw      & \gate{X^{m_0}}   & \ctrl{1} & \gate{H}  & \meter{} &  \cwbend{-2}   & m_2    \\
      t2   \qw & \qw        & \qw       & \qw & \qw      & \qw         & \gate{U_2}       & \qw        & \qw &  \qw &\qw  \\
                \lstick[2]{C}    a_3   \qw & \ctrl{}{}  & \qw       & \qw & \qw      & \gate{X^{m_0}}   & \ctrl{1} & \gate{H}  & \meter{} &  \cwbend{-2}   & m_3    \\
      t3   \qw & \qw        & \qw       & \qw & \qw      & \qw         & \gate{U_3}       & \qw        & \qw &  \qw &\qw  
        %\arrow[from=2-3,to=1-3,red,yshift=0.3cm,xshift=0.3cm,thick]{}
      %  \arrow[from=2-2,to=1-2,squiggly,dash,line width=0.1mm]{}
         \arrow[from=2-2,to=4-2,squiggly,dash,line width=0.1mm]{}
         \arrow[from=7-2,to=4-2,squiggly,dash,line width=0.1mm]{}
               %  \arrow[from=5-6,to=3-6,equal,dash]{}
    \end{quantikz}
    }
    $\equiv$
    \begin{quantikz}
 c~ & \ctrl{3}\gategroup[4,steps=1,style={dashed,inner sep=6pt}]{Dfanout}   & \qw  \\
 t_1  & \gate{U_1}   & \qw \\ 
  t_2  & \gate{U_2}   & \qw \\ 
  t_3  & \gate{U_3}   & \qw 
\end{quantikz}
\end{center}
\caption{Distributed fanout operation with single control qubit (on $A'$) for multiple target qubits (one on $A$, one on $B$ and one on $C$) - all target qubits on different nodes from the control qubit. The right hand side shows the notation we will use for distributed fanout throughout the paper. Note that $m_i \in \{0,1\}$. Here, the qubits $a_0,a_1,a_2$ and $a_3$ form a distributed GHZ state.}
\label{dfanout}
\end{figure*}
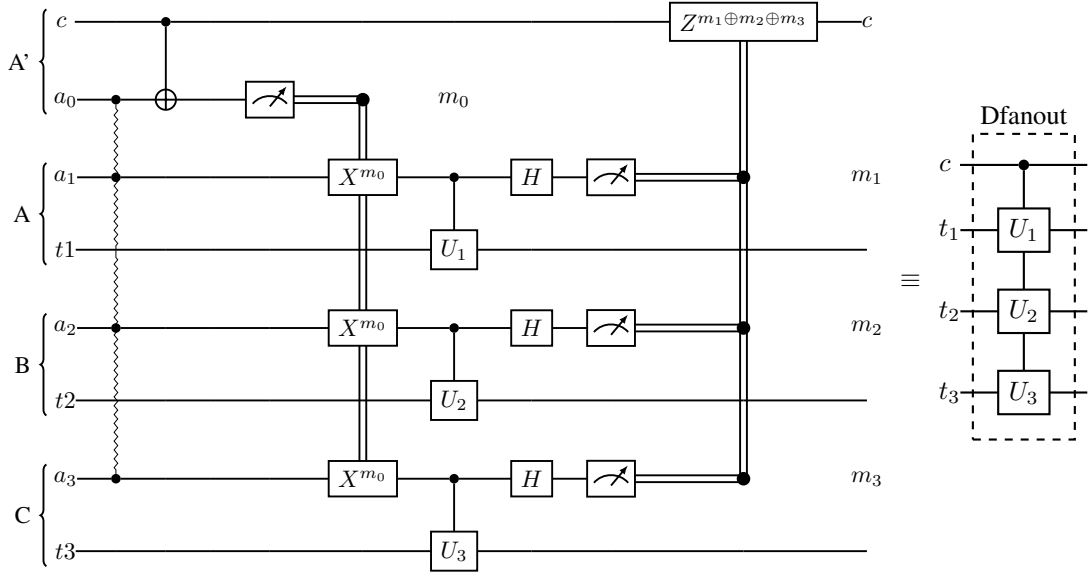

By a distributed fanout  operation, we refer to an operation where the same control qubit is used for multiple target qubits, on different nodes. This can arise from the structure of a quantum circuit itself, or from a control-$U$ operation where the unitary $U$ spans multiple qubits and is decomposed into operations executed on qubits distributed on different nodes~\cite{Yimsiriwattana:2004xhy,loke2023distributed,PhysRevA.107.L060601}. For example, the quantum circuit such as:
    \begin{center}
          \begin{quantikz}
     c~     & \ctrl{1} & \ctrl{2}  & \ctrl{3}  & \qw       \\
     t_1     & \gate{U_1} &  \qw  & \qw  & \qw      \\
     t_2     & \qw        &  \gate{U_2}   & \qw   & \qw    \\
     t_3      & \qw     &  \qw  & \gate{U_3}  & \qw             \\
        %\arrow[from=2-3,to=1-3,red,yshift=0.3cm,xshift=0.3cm,thick]{}
      %  \arrow[from=2-2,to=1-2,squiggly,dash,line width=0.1mm]{}
       %  \arrow[from=3-2,to=2-2,squiggly,dash,line width=0.1mm]{}
    \end{quantikz}
    ~$\equiv$~
    \begin{quantikz}
 c~ & \ctrl{3}\gategroup[4,steps=1,style={dashed,inner sep=6pt}]{fanout}   & \qw  \\
 t_1  & \gate{U_1}   & \qw \\ 
  t_2  & \gate{U_2}   & \qw \\ 
  t_3  & \gate{U_3}   & \qw 
\end{quantikz}
    \end{center}
i.e., a {\em multitarget control-$U$},   where $U_1$ acts on qubit~$t1$, $U_2$ acts on qubit~$t2$, and $U_3$ acts on qubit~$t3$, each qubit on a different node, will result in the distributed operation as shown in Figure~\ref{dfanout}; note that the wavy line in the figure connecting the four (communication) qubits $a_0,a_1,a_2,a_3$  represents a distributed 4-qubit GHZ state of the form $\frac{1}{\sqrt{2}}(\ket{0 0 0 0}+\ket{1111})_{a_0 a_1 a_2 a_3}$, over four nodes $A'$, $A$, $B$ and $C$. 
In the rest of the paper, we particularly focus on the case where  $U_i = X$, unless otherwise noted.

\section{Transversal Non-Local Fanout}
 
A transversal operation can be defined as follows.
Let there be $m$ code blocks, each consisting of $n$ physical qubits. A quantum operation $U$ is transversal if it can be expressed as $U=\bigotimes_{i=1}^{n} U_i$,
where each $U_i$ acts only on the $i$-th physical qubits of the code blocks and does not couple two qubits within the same code block.

From~\cite{Stack2026Transversal}, we see that a logical CNOT  between two logical qubits is transversal, when the logical qubits are encoded using the Bivariate-Bicycle (BB) qLDPC code; multiple logical CNOT operations are performed simultaneously.
Following ~\cite{transversalfanout}, because the logical fanout decomposes into CNOTs between the same code blocks, and because each of those logical CNOTs is transversal, their product can be reordered into a set of  independent physical fanout operations, one per physical coordinate; a logical fanout gadget is transversal if it can be composed using transversal logical CNOT gadgets. This comes from~\cite{Gottesman}, where ``a product of transversal gate gadgets is a transversal gate gadget'', and since, a logical fanout is a product of (transversal) CNOT operations, by definition.
Using the BB encoding $[[n,k,d]]$ of logical qubits, with operations on a block of $n$ qubits, we effectively perform $k$  independent logical fanout operations simultaneously.

Using the $[[n,k,d]]$ $BB$ encoding, where the logical CNOT is transversal,  i.e., between two blocks $C$ with qubits $c_i$ and $T$ with qubits $t_i$, denoted by $\overline{CNOT}(C;T) = \Pi_{i\in \{1,...,n\}}CNOT(c_i;t_{i})$,
consider  one control block $C = \{c_1,...,c_n\}$ with $n$ qubits,  and $m$ target blocks $T_i$, $i \in \{1,...,m\}$, each with qubits $T_{i}=\{t_{i1},...,t_{in}\}$, where block $C$ encodes $k$ logical qubits $\{ \bar{c}^{(l)} ~|~ l \in \{1,...,k\}\}$ and each of the $m$ blocks, $T_i$, encodes $k$ logical qubits  $\{ \bar{t}^{(l)}_{i} ~|~ l \in \{1,...,k\} \}$. 
Suppose we wanted to perform the following operation:
\begin{align*}
 & \overline{FANOUT}(C;T_1,...,T_m) \\
:= ~~ & \bigotimes_{l \in \{1,...,k\}}  \overline{FANOUT}_L (\bar{c}^{(l)};\bar{t}^{(l)}_{1},...,\bar{t}^{(l)}_{m}) 
\end{align*} 
where $\overline{FANOUT}_L$   denotes the logical fanout acting on the corresponding $l$-th encoded logical qubit, $1 \leq l \leq k$. We do this by performing $n$ independent physical fanout operations, one for each coordinate (corresponding to a physical qubit) of the code block:
\begin{align*}
 &   \bigotimes_{i \in \{1,...,n\}} FANOUT(c_i;t_{1i},...,t_{mi})  
 \end{align*} 
% = &\;\;\;\;     \overline{FANOUT}%(\bar{c}^{(1)};\bar{t}^{(1)}_{1},...,\bar{t}^{(1)}_{m}) \\
% & \otimes  \overline{FANOUT}(\bar{c}^{(2)};\bar{t}^{(2)}_{1},...,\bar{t}^{(2)}_{m})   \\
% & \otimes  ... \\
% & \otimes  \overline{FANOUT} (\bar{c}^{(k)};\bar{t}^{(k)}_{1},...,\bar{t}^{(k)}_{m}) \\
 where $FANOUT$ denotes a physical fanout operation, each physical qubit $c_i$ fans out as control to the corresponding physical qubits in every target block~\cite{transversalfanout}. Note that the above applies to non-local CNOTs and non-local FANOUTs, e.g., with one block per node (or $k$ logical qubits per node).
 
 %\footnote{We explain why this is so, as follows:
%\begin{align*}
 % &   \Pi_{i \in \{1,...,n\}} FANOUT(c_i;t_{1i},...,t_{mi})  \\
 %= & \Pi_{i \in \{1,...,n\}}  \Pi_{j\in \{1,...,m\}}CNOT(c_i;t_{ji}) \\
% & \mbox{(from definition of $FANOUT$)} \\
% = & \Pi_{j \in \{1,...,m\}}  \Pi_{i\in \{1,...,n\}}CNOT(c_i;t_{ji}) \\
 %& \mbox{(the required $CNOT$s commute)} \\
 %= & \Pi_{j \in \{1,...,m\}}  \overline{CNOT}(C;T_j) \\
 % & \mbox{(from transversal $CNOT$s between the control block $C$} \\
 %& \mbox{and the target block $T_j$)} \\ 
 %=  & \overline{FANOUT}(C;T_1,...,T_m)  
 %\end{align*}

%From another perspective, since a transversal CNOT between code blocks acts on  $k$ target logical qubits simultaneously, a transversal fanout among code blocks acts on  $k$ groups of target logical qubits simultaneously:
% \begin{align*}
 %  & \overline{FANOUT}(C;T_1,...,T_m)  \\
  %  = & \Pi_{j \in \{1,...,m\}}  \overline{CNOT}(C;T_j) \\
  %= & \Pi_{j \in \{1,...,m\}}  \Pi_{l\in \{1,...,k\}} \overline{CNOT}_L(\bar{c}^{(l)};\bar{t}_j^{(l)}) \\
  %= & \Pi_{l\in \{1,...,k\}} \Pi_{j \in \{1,...,m\}}  \overline{CNOT}_L(\bar{c}^{(l)};\bar{t}_j^{(l)}) \\
 %& \mbox{(the required $\overline{CNOT}_L$s commute)} \\
%=  & \Pi_{l \in \{1,...,k\}}  \overline{FANOUT}_L(\bar{c}^{(l)};\bar{t}^{(l)}_{1},...,\bar{t}^{(l)}_{m}) \\
 %& \mbox{(from definition of $\overline{FANOUT}_L$)}  \\
 %=  & \bigotimes_{l \in \{1,...,k\}}  \overline{FANOUT}_L(\bar{c}^{(l)};\bar{t}^{(l)}_{1},...,\bar{t}^{(l)}_{m})
%\end{align*} 
%}

\section{Simulation analysis with Different Encodings}
% Add the following packages to your preamble if they are not already present:
% \usepackage{graphicx}
% \usepackage{amsmath}
% The PNG files should be in the same directory as the main .tex file,
% or change the paths in the \includegraphics commands accordingly.

\begin{figure}[ht!]
\centering
  \includegraphics[width=0.5\textwidth,height=6.5cm]{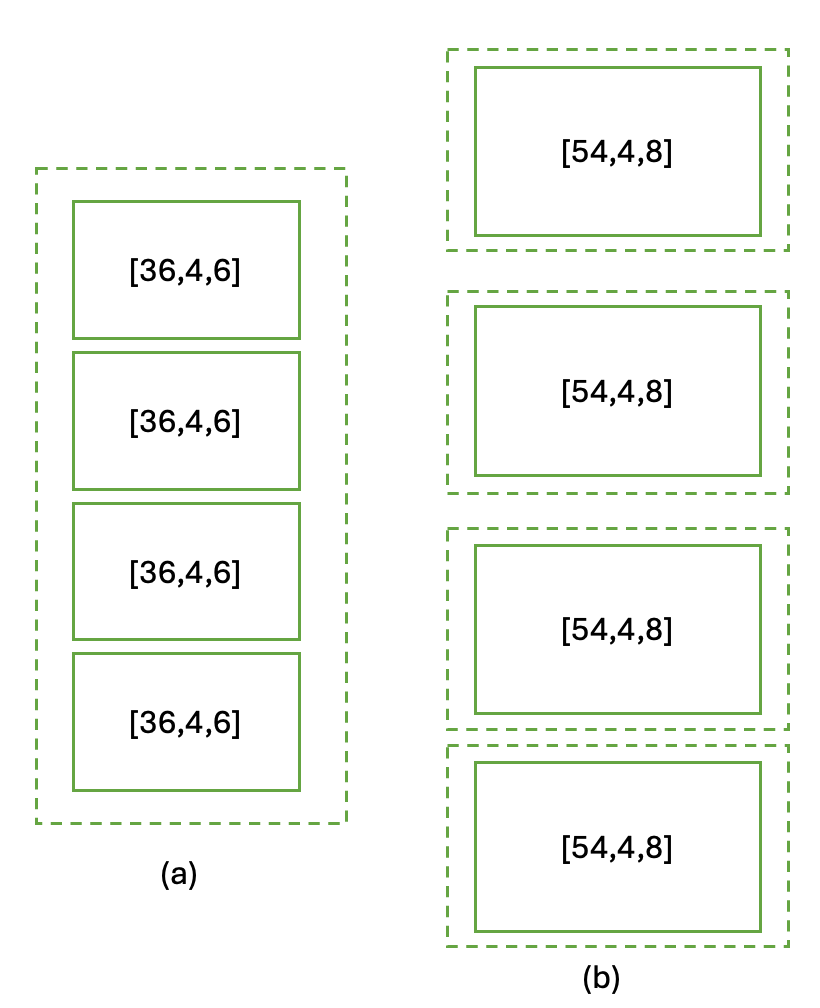}      
  \caption{Configurations used in our experiments - four nodes: (a) a configuration with four blocks on one node (all operations local) and (b) a configuration with four nodes, one block per node, but with a higher distance code.}
\label{fig:config}
\end{figure}

We consider a 4-node configuration with one code block per node in most of our analysis below; for comparison, we have a 4 blocks on one node configuration. With all four blocks, we have a total of 16 logical qubits  involved in 4 concurrent distributed  transversal fanout operations (each involving 4 logical qubits, one from each node). Figure~\ref{fig:config} illustrates the  configurations used in our experiments.

We used   the Transversal Multiple Code Block Simulator (TMCBS)
library code from\cite{Stack2026Transversal} (developed for simulating non-local transversal CNOTs), which uses stim,\footnote{https://github.com/quantumlib/stim} and  we use sinter\footnote{https://pypi.org/project/sinter/}  for Monte Carlo sampling.

\begin{figure}[ht!]
\centering
\includegraphics[width=0.5\textwidth]{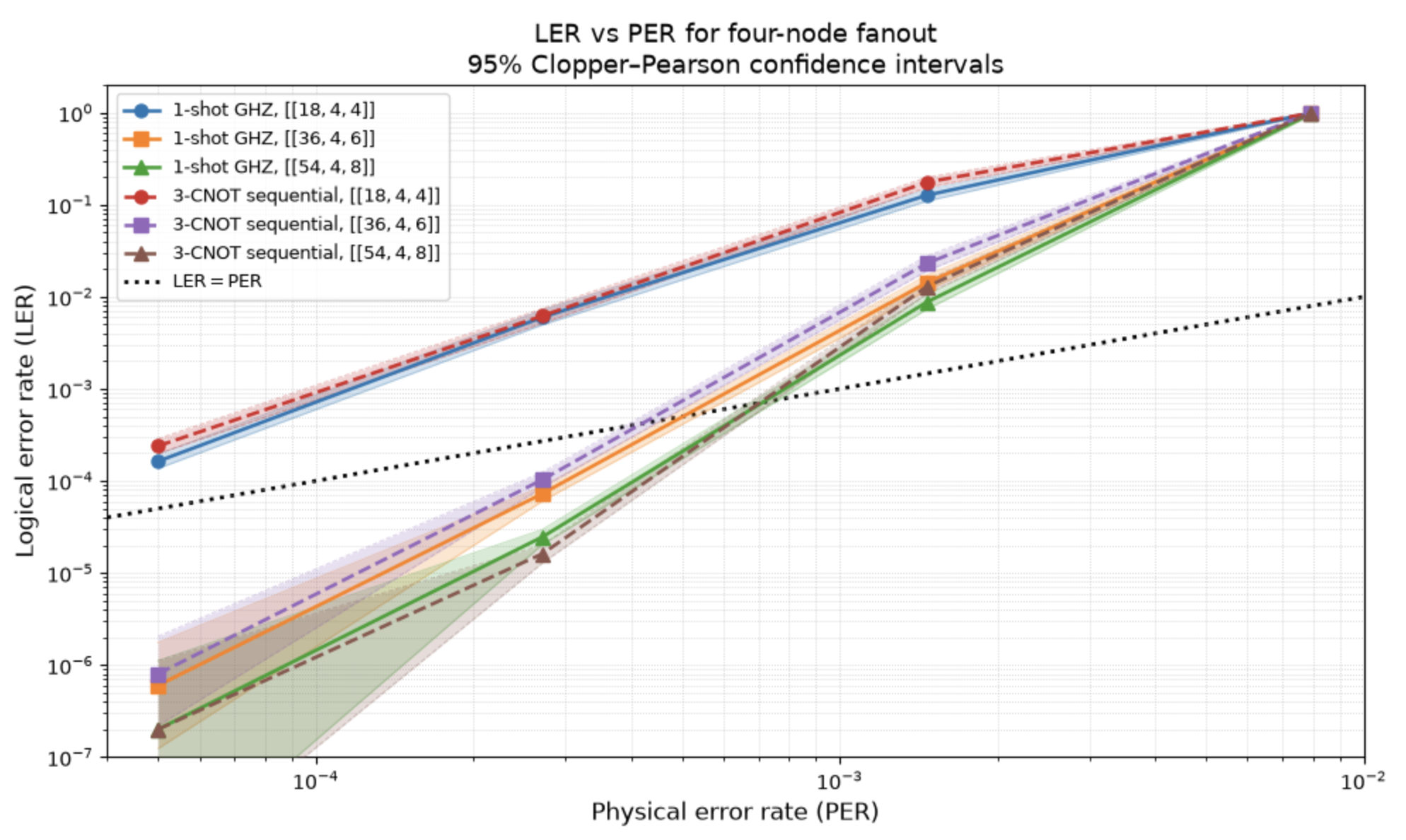}
\caption{Two versions of four node fanout for three different encodings: logical error rate (LER) versus physical error rate (PER), with shaded exact two-sided 95\% Clopper--Pearson confidence intervals. For varying PER=$p$, we set $p_{ebit}=p_{ghz}=10p$. 
Table~\ref{tab:combined-fanout-ler-per} in the Appendix shows the figures plotted here.}
\label{fig:combined-ler-per} 
\end{figure}

Figure~\ref{fig:combined-ler-per}   shows the graph of LERs vs PERs, using $p_{ebit}=p_{ghz}=10p$, for PER $p$, using three different BB encodings $[[18,4,4]]$,  $[[36,4,6]]$, and $[[54,4,8]]$, for two simulated realizations of a distributed (or non-local) fanout, one using a one shot approach to create distributed GHZ states (1shotghz-fanout), and one using   a sequence of 3 (non-local) CNOTS (since there are four blocks), equivalent to the fanout (3cnot-seq). 
We construct (blocks of) GHZ states for 1shotghz-fanout   via a transversal protocol: among corresponding qubits in the communication qubit blocks on each node, perform the Hadamard $H$ on one ``root'' (control) qubit, and then perform non-local CNOT between the ``root'' and each of the other qubits; we count the entire process as taking one simulation step, to simulate the one shot approach to creating GHZ states (one shot GHZ approaches are possible, e.g.~\cite{6mqy-sd3d,Ainley:2024bdu}). The respective noise models and   
syndrome extraction are similar to ~\cite{transversalfanout} and is summarised here in Table~\ref{tab:syndrome-extraction}. 
For each data point, we also show error bars corresponding to 95\% confidence intervals.

\begin{table*}[ht!]
\centering
\caption{Syndrome-extraction schedule for  the different distributed  fanout 
  implementations, and the distributed CNOTs (dCNOT)s they are constituted by.  For each,   initial extraction consists of one 
first-pass round followed by three repeated rounds. For 
\texttt{3cnot-seq}, an additional extraction round is performed 
between successive non-local CNOT operations; this isolates successive logical operations and provide the same fault-tolerance boundary used by the TMCBS non-local-CNOT protocol. No such intermediate extraction is required in the GHZ-fanout circuit because the fanout is treated as one logical operation.
DEPOLARIZE2 in Stim is a two-qubit depolarizing noise channel, with identity with probability $1-p_{ebit}$, and $p_{ebit}$, otherwise, i.e.
$Pr(II) =1-p_{ebit}, Pr(P) =\frac{p_{ebit}}{15}$, for $P\in\{IX,IY,IZ,XI,\ldots,ZZ\}$, and 
$\mathcal{E}_{\mathrm{GHZ}}(p_{\mathrm{ghz}}) =\prod_q \mathcal{D}_1(q,p_{\mathrm{ghz}}/2)$, where the implementation applies independent $\mathcal{D}_1$ (DEPOLARIZE1) channels to the (distributed) GHZ qubits (where DEPOLARIZE1 is Stim’s single-qubit depolarizing channel) giving identity with probability $1-p_{\mathrm{ghz}}/2$ and $(p_{\mathrm{ghz}}/2)/3$ for $X$, $Y$ and $Z$ error, in this case. Tesseract is the decoder used. }
\label{tab:syndrome-extraction}
\begin{tabular}{lcccc}
\hline
\textbf{Circuit} &
\textbf{Before non-local operation} &
\textbf{Between operations} &
\textbf{After operation}  &
\textbf{Non-local noise model} \\\\
\hline

\texttt{3cnot-seq}
& 4 rounds
& 1 round after each non-local CNOT
& 3 rounds 
& DEPOLARIZE2 per dCNOT \\

\texttt{1shotghz-fanout}
& 4 rounds
& ---
& 3 rounds 
& $\mathcal{E}_{\mathrm{GHZ}}$ per dGHZ \\
\hline
\end{tabular}
\end{table*}

Table~\ref{tab:combined-fanout-ler-per} in the Appendix shows the figures plotted in Figure~\ref{fig:combined-ler-per}.
It can be seen that even when larger blocks are used with more physical qubits and higher distance encodings, even though more ebits are required for the same number of logical qubits, the higher distances can compensate for and in fact enables a lower LER for the fanout operation. In particular, $[[54,4,8]]$ achieves a lower LER than $[[36,4,6]]$ even with $54/36=1.5$ times more ebits (or more distributed GHZ states), and in contrast to $[[18,4,4]]$, achieves $LER < PER$ with more than double the number of ebits used for each $dCNOT$.

\begin{figure}[ht!]
\centering
\includegraphics[width=0.5\textwidth]{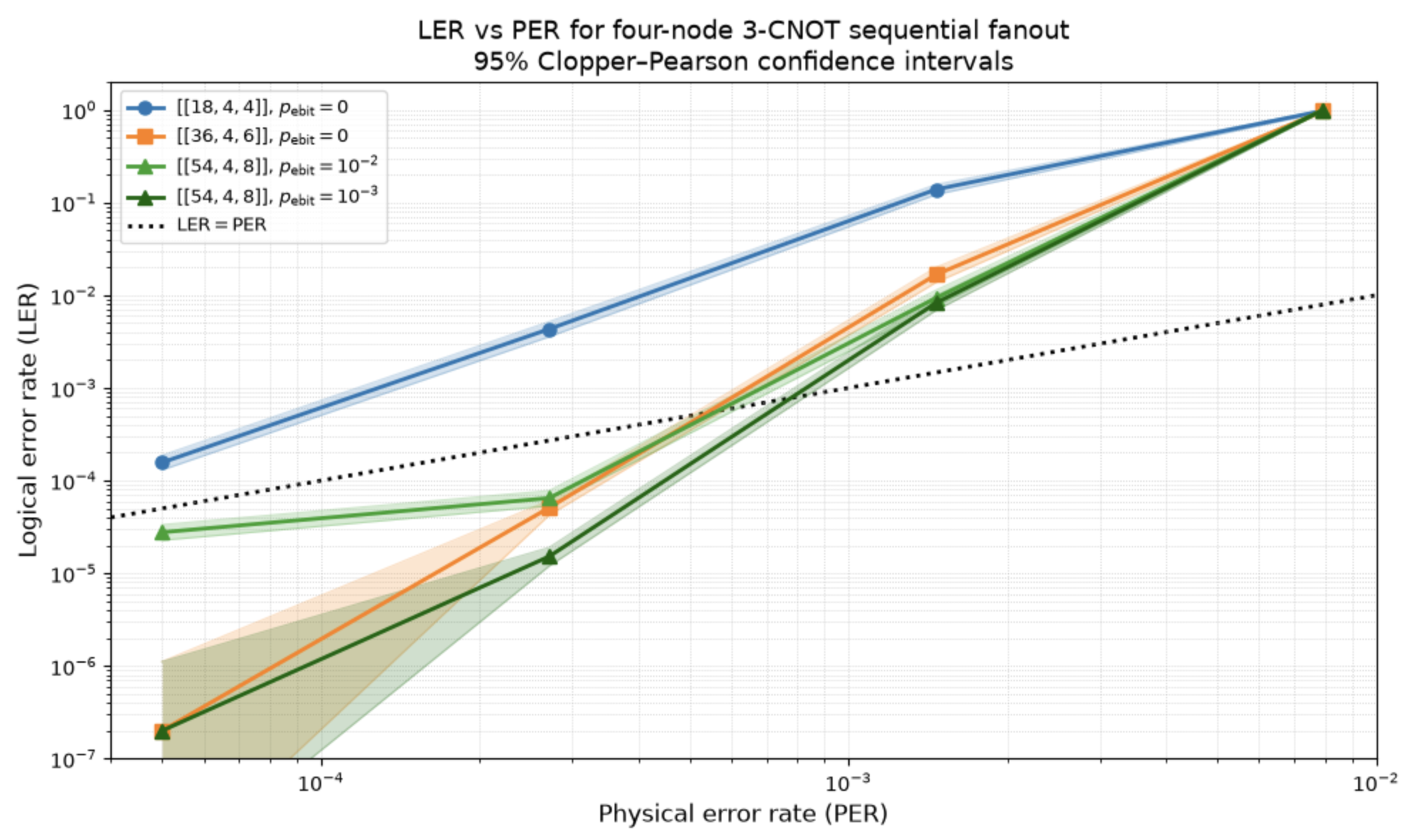}
\caption{Four node 3-CNOT sequential fanout: logical error rate (LER) versus physical error rate (PER), with shaded exact two-sided 95\% Clopper--Pearson confidence intervals. For varying PER=$p$, we used a fixed  $p_{ebit}$. Table~\ref{tab:cnotseq-ler-per} in the Appendix shows the data for this graph.}
\label{fig:cnotseq-ler-per}
\end{figure}

 Figure~\ref{fig:cnotseq-ler-per} shows a similar trend with fixed $p_{ebit}$ where the distributed fanouts are implemented using  sequences of non-local transversal CNOTs (3 $dCNOTs$ for each fanout since there are 4 nodes). We introduced the two cases of $p_{ebit}=0$ in the simulation with the $[[18,4,4]]$ and $[[36,4,6]]$ encodings for comparison - effectively modelling all qubits (or all blocks) and so, all operations on the same node (note that there are extra operations involving local ebits which would not be needed compared to a totally local fanout using all local CNOTs - hence, this can be viewed as an upper bound or higher estimate of a completely local fanout). One observation is that the $[[54,4,8]]$ encoding, when $p_{ebit}=10^2$ (quite a high error rate), just managed to keep the $LER$ below the $PER$ for low enough $p$, but when  with a lower $p_{ebit}=10^3$, the  $[[54,4,8]]$  encoding comfortably pushes $LER$ well below the $PER$. This means that high network error rates (modelled by the $p_{ebit}$) can reduce the fault tolerance of the encoding substantially. Also, at $p_{ebit}=10^3$, the  $[[54,4,8]]$ encoding achieves a similar $LER$ as the entirely local operation with  $[[36,4,6]]$ blocks, that is, the  distributed fanout involving 16 logical qubits over 4  nodes, each node with 4 logical qubits in a  $[[54,4,8]]$ block (in Figure~\ref{fig:config}(b)), has equivalent $LER$ as the local fanout involving 16 logical qubits encoded in 4 blocks of $[[36,4,6]]$ on one node (in Figure~\ref{fig:config}(a)), at low  enough $p_{ebit}$; in terms of $LER$, the additional 4x54-4x36=216-144=72 physical computation/data qubits to increase the distance from $6$ to $8$ compensated fully for the non-local operations - in other words, instead of a node with 144  physical data qubits, one can equivalently (in terms of $LER$) use 4 nodes of $54$ physical data qubits, useful if no such 144  physical data qubits machine is available.

\begin{figure}[ht!]
\centering
\includegraphics[width=0.5\textwidth]{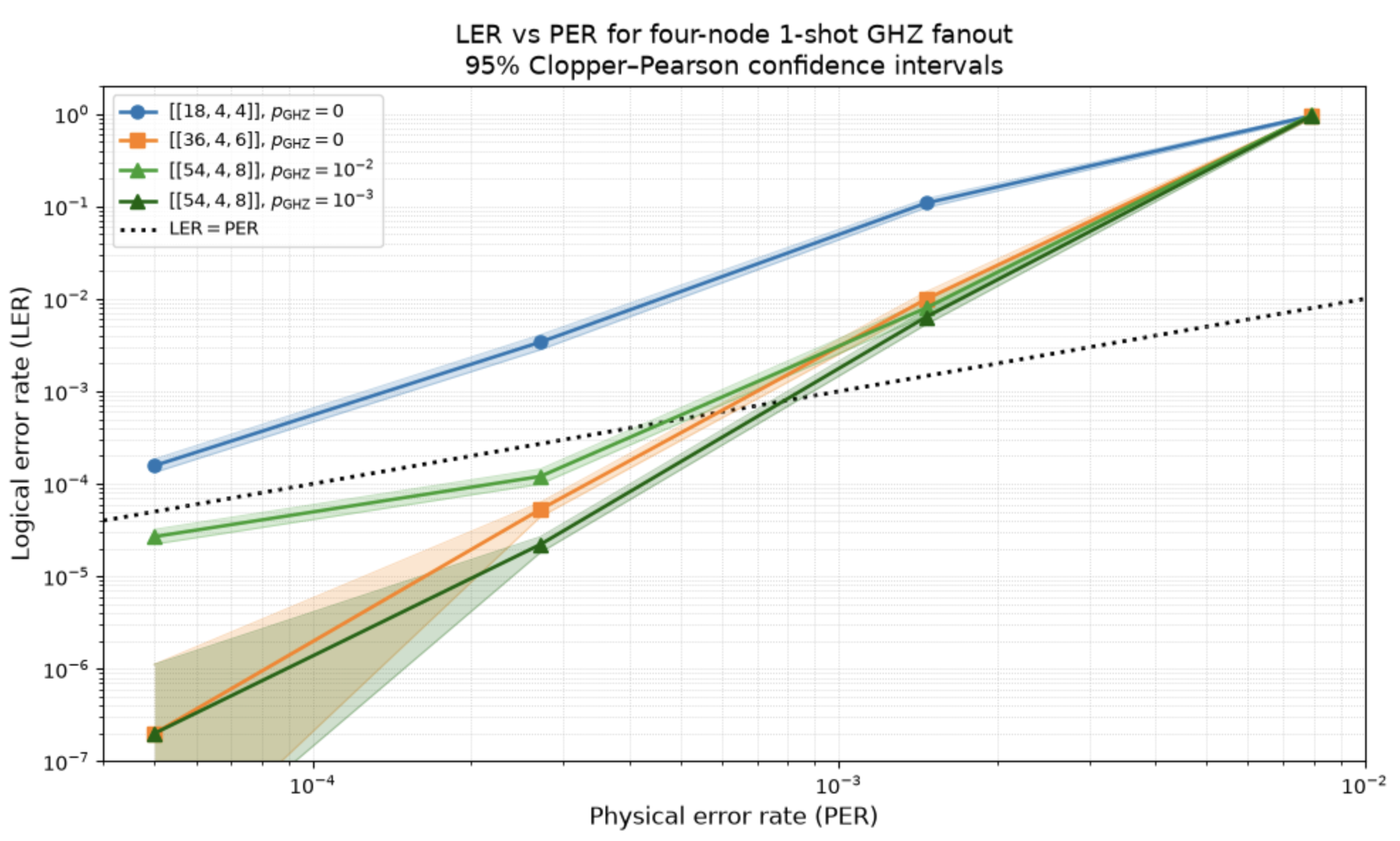}
\caption{Four node 1-shot GHZ fanout: logical error rate (LER) versus physical error rate (PER), with shaded exact two-sided 95\% Clopper--Pearson confidence intervals. For varying PER=$p$, we used a fixed  $p_{ghz}$.  Table~\ref{tab:ghz-ler-per} in the Appendix shows the data for this graph.}
\label{fig:ghz-ler-per}
\end{figure}

For a distributed fanout implemented transversally using distributed GHZ states,
 Figure~\ref{fig:ghz-ler-per} shows a similar trend with fixed $p_{ghz}$ at $10^2$ and $10^3$. The higher distance compensated for the larger number of  distributed GHZ states used for the $[[54,4,8]]$ blocks, especially at low  enough $p_{ghz}$.  A similar equivalence in terms of $LER$ is found between the fanouts with distributed 16 logical qubits and the entirely local operations with four $[[36,4,6]]$ blocks on one node (modelled by setting $p_{ghz}=0$).
(A deeper comparison between distributed fanouts using dCNOT sequences and dGHZ states is in~\cite{transversalfanout}, and is not done in this paper.)
 
\section{Implications on Node Architecture Design}
\begin{figure}[h!]
\centering
  \includegraphics[width=0.5\textwidth]{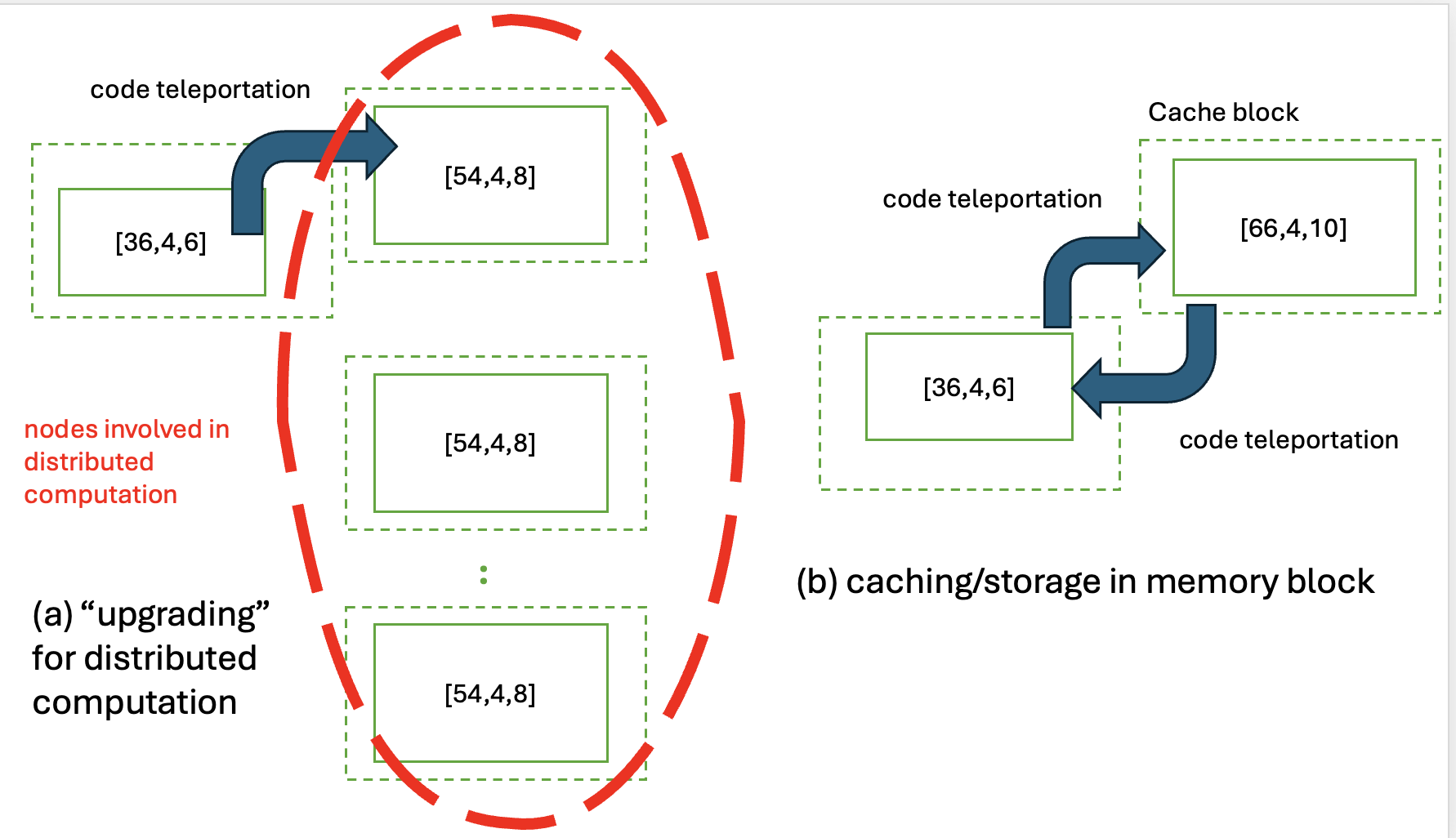}      
  \caption{Two uses of ``upgrading'' logical qubit encodings: (a) to increase ``protection'' before distributed operations, and (b) to temporarily cache qubits while waiting for distributed operations, e.g., delays in obtaining ebits, etc; how much (higher) the distance encoding required may depend on the required wait time.}
\label{fig:tele}
\end{figure}

The above observations suggest that code teleportation,  used to convert from one encoding to another, e.g., \cite{6675561},   also used to bring logical qubit states from memory qubits into processor qubits~\cite{Brun:2015asy,cain2026shor}, or code switching (e.g., \cite{Liang2026UltraHighDistance,Heussen2025efficientfault}) could be used to  ``upgrade'' a block of logical qubits, i.e., to convert from a code $[[n,k,d]]$ to $[[n',k,d']]$, where $n' > n$ and $d' > d$. This can be done, for example, to:
(i) improve code distance and error protection for the logical qubits before they get involve in distributed computations; and/or
(ii) cache logical quantum qubit states, e.g., if there are network delays, by moving the qubits into higher distance blocks, as illustrated in Figure~\ref{fig:tele}.

\begin{figure}[ht!]
\centering
  \includegraphics[width=0.5\textwidth]{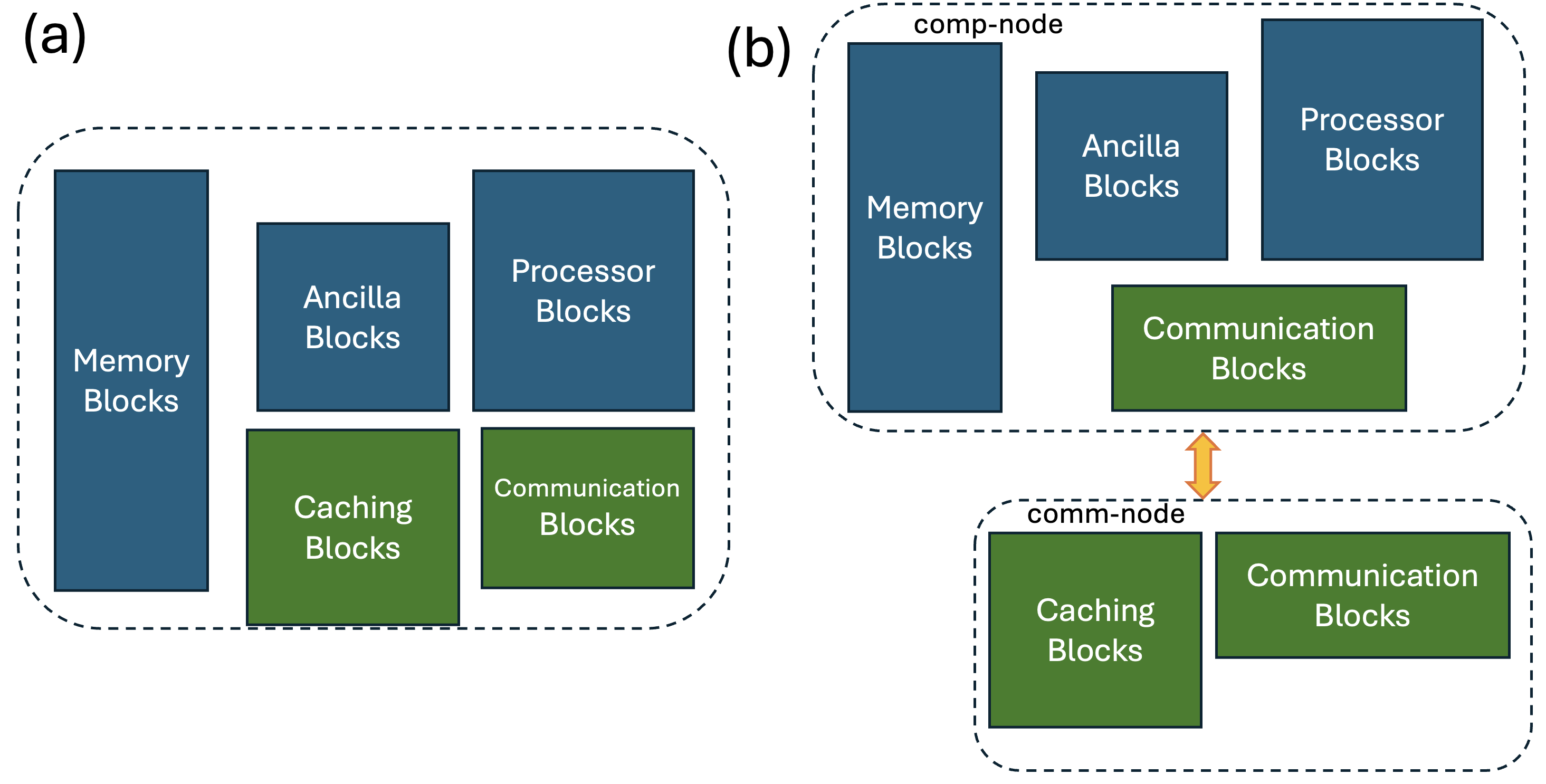}      
  \caption{Communication and cache  blocks of qubits as (a) part of a node, or (b) on their own node (usable by comp-nodes).}
\label{fig:node-arch}
\end{figure}

\begin{figure}[ht!]
\centering
\includegraphics[width=0.5\textwidth]{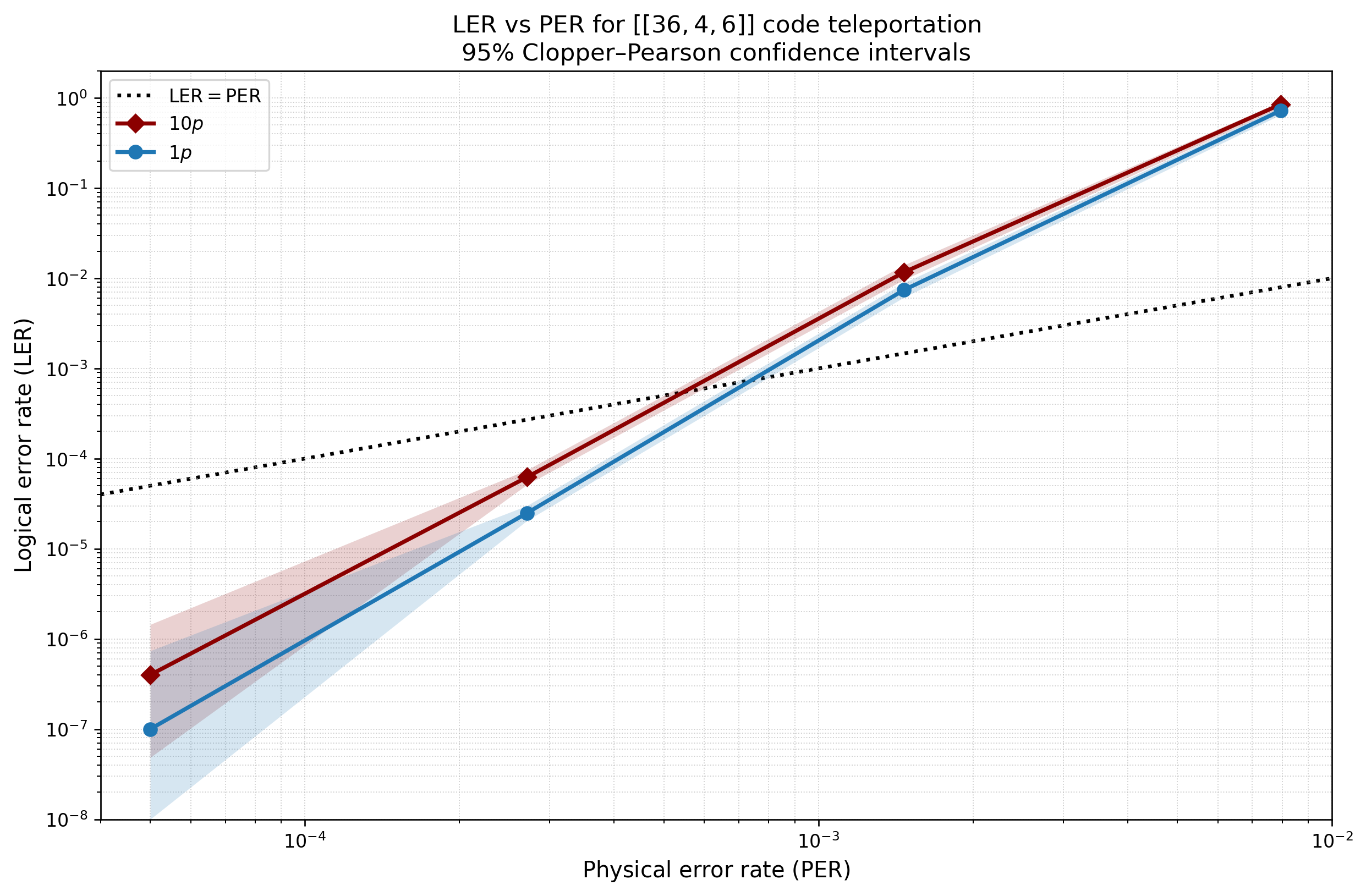}
\caption{LER for fault-tolerant teleportation with $[[36,4,6]]$ blocks across two nodes with $p_{ebit}=10p$ and $p_{ebit}=p$. Table~\ref{tab:36-4-6-teleportation-ler-per} in the Appendix shows the data for this graph.}
\label{fig:tele-ler-per}
\end{figure}

The use of such caching encoding blocks of qubits might call for an architecture of nodes involved in distributed quantum computations to have, beyond blocks of memory, ancilla, and processor qubits (as mentioned in ~\cite{Brun:2015asy,cain2026shor}), blocks of qubits for communication (holding ebits), and cache qubits for the two purposes above, as illustrated in Figure~\ref{fig:node-arch}.  It could be that the blocks of communication qubits and cache qubits (additional to memory, ancilla and processor qubit blocks) are hosted on a separate node (call these {\em comm-nodes}), to be usable by different nodes (the {\em comp-nodes}), as illustrated in Figure~\ref{fig:node-arch}(b); for this architecture, logical qubits from a node can be teleported into the remote cache on a comm-node - the cache may have a higher distance encoding then the comp-nodes for the logical qubits (but assuming a high quality classical and quantum link between the comp-node and the comm-node, e.g., within the same locality), depending on the required wait time.  
Such code teleportation between encoding blocks  on different nodes can utilise logical Bell pairs which can be efficiently formed (e.g., the work in~\cite{tz8f-6xwk}, which needs to be adapted for BB codes); for example, imagining three blocks of logical qubits, one on the comp-node (holding the logical data qubits), and two hosting logical Bell pairs hosted across the comp-node and comm-node,  the remote code teleportation then can be done with local joint logical qubit measurements on the comp-node with corrections on the comm-node (as in a typical quantum teleportation protocol) but on logical qubits and using logical operations - for blocks with $[[n,k,d]]$ encoding, concurrent teleportation of $k$ logical qubits can be carried out.   Several possibilities exist - e.g., one is direct (remote) teleportation  from a  $[[n,k,d]]$ block on the comp-node to one with $[[n',k,d']]$ on the comm-node, with $n' > n$ and $d' > d$, or teleportation first among equal encoding blocks and then locally ``upgrade'' (as needed).
Assuming a high quality link between the comm-node and comp-node (e.g., $p_{ebit}=p$), teleportation can have a relatively low LER, indicated by Figure~\ref{fig:tele-ler-per}, among equal sized blocks, e.g.,  at least one or several orders of magnitude lower than the LER of the distributed transversal fanout itself.

Local operations could work with a ``weaker protection'' encoding for the logical qubits, requiring fewer physical qubits per logical qubit, while distributed operations may require a higher encoding to increase chances of success. However, additional resources such as fault-tolerant operations are required for the ``upgrading'' (or for ``downgrading'' later if required), so that a further cost-benefit analysis is required. 

\section{Conclusion and Future Work}
We investigated distributed   fanout gadgets using different BB encodings of logical qubits.  We observed that higher distance encodings with more physical qubits (even though using more ebits or more distributed GHZ states, due to distributed transversal operations), can help deal with logical errors to achieve $LER <PER$, but also comparable with entirely local operations (with lower distances), at low enough $p_{ebit}$ and $p_{ghz}$. This suggests a node architecture to consider ``upgrading'' encoding to maintain quantum information before they are involved in more error-prone distributed operations, or to cache quantum information in case of network delays.   Further work is needed to estimate the trade-off in fault-tolerant code switching versus the benefits. Also, similar studies can be done with other distributed quantum circuits, e.g., to see when such ``upgrading'' is useful, to  study under what network conditions caching is useful, and to understand what actual encodings to use for a given circuit and a given network quality, in order to achieve a target LER. While our results have used small sized blocks, much larger encoding blocks and larger circuits can be analyzed with higher code distances, where, hopefully, a similar trend can be observed - this is an avenue for future work.

\section*{Acknowledgements}
The author acknowledges the use of ChatGPT-6  to assist in generating the Python code for plotting the graphs in Figures~\ref{fig:combined-ler-per}, \ref{fig:cnotseq-ler-per},  \ref{fig:ghz-ler-per}, and \ref{fig:tele-ler-per}. The author verified all algorithmic logic and executed the code, and takes full responsibility for the  results.
\bibliographystyle{plain}
\bibliography{refs}

\section*{Appendix}

\begin{table*}[p]
\centering
\caption{Data for the 4-node fanout LER-versus-PER graph. The logical error rate is $o/s$, with $o$ observed logical errors in $s$ shots. Confidence limits are exact two-sided 95\% Clopper--Pearson intervals. We set $p_{ebit}=p_{ghz}=10p$, $p_{ghz}=10p$}
\label{tab:combined-fanout-ler-per}
\scriptsize
\setlength{\tabcolsep}{4pt}
\resizebox{\textwidth}{!}{%
\begin{tabular}{lllr r r r r}
\hline
Method & Code  & PER & Shots & Errors & LER & 95\% CI for LER \\
\hline
1-shot GHZ & $[[18,4,4]]$   & $5.00\times 10^{-5}$ & 796,262 & 130 & $1.63\times 10^{-4}$ & [$1.36\times 10^{-4}$, $1.94\times 10^{-4}$] \\
1-shot GHZ & $[[18,4,4]]$  & $2.71\times 10^{-4}$ & 18,978 & 114 & $6.01\times 10^{-3}$ & [$4.96\times 10^{-3}$, $7.21\times 10^{-3}$] \\
1-shot GHZ & $[[18,4,4]]$   & $1.47\times 10^{-3}$ & 1,314 & 167 & $1.27\times 10^{-1}$ & [$1.10\times 10^{-1}$, $1.46\times 10^{-1}$] \\
1-shot GHZ & $[[18,4,4]]$   & $7.94\times 10^{-3}$ & 137 & 136 & $9.93\times 10^{-1}$ & [$9.60\times 10^{-1}$, $1.00\times 10^{0}$] \\
\hline
1-shot GHZ & $[[36,4,6]]$  & $5.00\times 10^{-5}$ & 5,000,000 & 3 & $6.00\times 10^{-7}$ & [$1.24\times 10^{-7}$, $1.75\times 10^{-6}$] \\
1-shot GHZ & $[[36,4,6]]$  & $2.71\times 10^{-4}$ & 1,647,206 & 120 & $7.29\times 10^{-5}$ & [$6.04\times 10^{-5}$, $8.71\times 10^{-5}$] \\
1-shot GHZ & $[[36,4,6]]$   & $1.47\times 10^{-3}$ & 7,842 & 112 & $1.43\times 10^{-2}$ & [$1.18\times 10^{-2}$, $1.72\times 10^{-2}$] \\
1-shot GHZ & $[[36,4,6]]$   & $7.94\times 10^{-3}$ & 118 & 116 & $9.83\times 10^{-1}$ & [$9.40\times 10^{-1}$, $9.98\times 10^{-1}$] \\
\hline
1-shot GHZ & $[[54,4,8]]$   & $5.00\times 10^{-5}$ & 5,000,000 & 1 & $2.00\times 10^{-7}$ & [$5.06\times 10^{-9}$, $1.11\times 10^{-6}$] \\
1-shot GHZ & $[[54,4,8]]$   & $2.71\times 10^{-4}$ & 4,576,102 & 113 & $2.47\times 10^{-5}$ & [$2.04\times 10^{-5}$, $2.97\times 10^{-5}$] \\
1-shot GHZ & $[[54,4,8]]$   & $1.47\times 10^{-3}$ & 14,570 & 128 & $8.79\times 10^{-3}$ & [$7.33\times 10^{-3}$, $1.04\times 10^{-2}$] \\
1-shot GHZ & $[[54,4,8]]$   & $7.94\times 10^{-3}$ & 110 & 108 & $9.82\times 10^{-1}$ & [$9.36\times 10^{-1}$, $9.98\times 10^{-1}$] \\
\hline
3-CNOT sequential & $[[18,4,4]]$   & $5.00\times 10^{-5}$ & 488,926 & 117 & $2.39\times 10^{-4}$ & [$1.98\times 10^{-4}$, $2.87\times 10^{-4}$] \\
3-CNOT sequential & $[[18,4,4]]$   & $2.71\times 10^{-4}$ & 18,978 & 118 & $6.22\times 10^{-3}$ & [$5.15\times 10^{-3}$, $7.44\times 10^{-3}$] \\
3-CNOT sequential & $[[18,4,4]]$   & $1.47\times 10^{-3}$ & 1,178 & 207 & $1.76\times 10^{-1}$ & [$1.54\times 10^{-1}$, $1.99\times 10^{-1}$] \\
3-CNOT sequential & $[[18,4,4]]$   & $7.94\times 10^{-3}$ & 121 & 120 & $9.92\times 10^{-1}$ & [$9.55\times 10^{-1}$, $1.00\times 10^{0}$] \\
\hline
3-CNOT sequential & $[[36,4,6]]$   & $5.00\times 10^{-5}$ & 5,000,000 & 4 & $8.00\times 10^{-7}$ & [$2.18\times 10^{-7}$, $2.05\times 10^{-6}$] \\
3-CNOT sequential & $[[36,4,6]]$   & $2.71\times 10^{-4}$ & 1,092,198 & 113 & $1.03\times 10^{-4}$ & [$8.53\times 10^{-5}$, $1.24\times 10^{-4}$] \\
3-CNOT sequential & $[[36,4,6]]$   & $1.47\times 10^{-3}$ & 4,834 & 112 & $2.32\times 10^{-2}$ & [$1.91\times 10^{-2}$, $2.78\times 10^{-2}$] \\
3-CNOT sequential & $[[36,4,6]]$   & $7.94\times 10^{-3}$ & 109 & 109 & $1.00\times 10^{0}$ & [$9.67\times 10^{-1}$, $1.00\times 10^{0}$] \\
\hline
3-CNOT sequential & $[[54,4,8]]$   & $5.00\times 10^{-5}$ & 5,000,000 & 1 & $2.00\times 10^{-7}$ & [$5.06\times 10^{-9}$, $1.11\times 10^{-6}$] \\
3-CNOT sequential & $[[54,4,8]]$   & $2.71\times 10^{-4}$ & 5,000,000 & 81 & $1.62\times 10^{-5}$ & [$1.29\times 10^{-5}$, $2.01\times 10^{-5}$] \\
3-CNOT sequential & $[[54,4,8]]$   & $1.47\times 10^{-3}$ & 10,370 & 134 & $1.29\times 10^{-2}$ & [$1.08\times 10^{-2}$, $1.53\times 10^{-2}$] \\
3-CNOT sequential & $[[54,4,8]]$   & $7.94\times 10^{-3}$ & 118 & 118 & $1.00\times 10^{0}$ & [$9.69\times 10^{-1}$, $1.00\times 10^{0}$] \\
\hline
\end{tabular}%
}
\end{table*}

\begin{table*}[p]
\centering
\caption{Data for the four-node 3-CNOT sequential fanout LER-versus-PER graph. The logical error rate is $k/n$, with $k$ observed logical errors in $n$ shots. Confidence limits are exact two-sided 95\% Clopper--Pearson intervals.}
\label{tab:cnotseq-ler-per}
\scriptsize
\setlength{\tabcolsep}{4pt}
\resizebox{\textwidth}{!}{%
\begin{tabular}{lllr r r r r}
\hline
Method & Code & Parameter & PER & Shots & Errors & LER & 95\% CI for LER \\
\hline
3-CNOT sequential & $[[18,4,4]]$ & $p_{\mathrm{ebit}}=0$
& $5.00\times 10^{-5}$ & 772,710 & 121
& $1.57\times 10^{-4}$
& [$1.30\times 10^{-4}$, $1.87\times 10^{-4}$] \\

3-CNOT sequential & $[[18,4,4]]$ & $p_{\mathrm{ebit}}=0$
& $2.71\times 10^{-4}$ & 26,146 & 113
& $4.32\times 10^{-3}$
& [$3.56\times 10^{-3}$, $5.19\times 10^{-3}$] \\

3-CNOT sequential & $[[18,4,4]]$ & $p_{\mathrm{ebit}}=0$
& $1.47\times 10^{-3}$ & 1,570 & 218
& $1.39\times 10^{-1}$
& [$1.22\times 10^{-1}$, $1.57\times 10^{-1}$] \\

3-CNOT sequential & $[[18,4,4]]$ & $p_{\mathrm{ebit}}=0$
& $7.94\times 10^{-3}$ & 151 & 148
& $9.80\times 10^{-1}$
& [$9.43\times 10^{-1}$, $9.96\times 10^{-1}$] \\
\hline
3-CNOT sequential & $[[36,4,6]]$ & $p_{\mathrm{ebit}}=0$ & $5.00\times 10^{-5}$ & 5,000,000 & 1 & $2.00\times 10^{-7}$ & [$5.06\times 10^{-9}$, $1.11\times 10^{-6}$] \\
3-CNOT sequential & $[[36,4,6]]$ & $p_{\mathrm{ebit}}=0$ & $2.71\times 10^{-4}$ & 2,370,150 & 121 & $5.11\times 10^{-5}$ & [$4.24\times 10^{-5}$, $6.10\times 10^{-5}$] \\
3-CNOT sequential & $[[36,4,6]]$ & $p_{\mathrm{ebit}}=0$ & $1.47\times 10^{-3}$ & 6,690 & 112 & $1.67\times 10^{-2}$ & [$1.38\times 10^{-2}$, $2.01\times 10^{-2}$] \\
3-CNOT sequential & $[[36,4,6]]$ & $p_{\mathrm{ebit}}=0$ & $7.94\times 10^{-3}$ & 138 & 135 & $9.78\times 10^{-1}$ & [$9.38\times 10^{-1}$, $9.95\times 10^{-1}$] \\
\hline
3-CNOT sequential & $[[54,4,8]]$ & $p_{\mathrm{ebit}}=10^{-2}$ & $5.00\times 10^{-5}$ & 3,863,142 & 107 & $2.77\times 10^{-5}$ & [$2.27\times 10^{-5}$, $3.35\times 10^{-5}$] \\
3-CNOT sequential & $[[54,4,8]]$ & $p_{\mathrm{ebit}}=10^{-2}$ & $2.71\times 10^{-4}$ & 1,713,203 & 111 & $6.48\times 10^{-5}$ & [$5.33\times 10^{-5}$, $7.80\times 10^{-5}$] \\
3-CNOT sequential & $[[54,4,8]]$ & $p_{\mathrm{ebit}}=10^{-2}$ & $1.47\times 10^{-3}$ & 10,754 & 101 & $9.39\times 10^{-3}$ & [$7.66\times 10^{-3}$, $1.14\times 10^{-2}$] \\
3-CNOT sequential & $[[54,4,8]]$ & $p_{\mathrm{ebit}}=10^{-2}$ & $7.94\times 10^{-3}$ & 105 & 105 & $1.00\times 10^{0}$ & [$9.65\times 10^{-1}$, $1.00\times 10^{0}$] \\
%3-CNOT sequential & $[[66,4,10]]$ & $p_{\mathrm{ebit}}=10^{-2}$ & $5.00\times 10^{-5}$ & 5,000,000 & 17 & $3.40\times 10^{-6}$ & [$1.98\times 10^{-6}$, $5.44\times 10^{-6}$] \\
%3-CNOT sequential & $[[66,4,10]]$ & $p_{\mathrm{ebit}}=10^{-2}$ & $2.71\times 10^{-4}$ & 3,381,094 & 104 & $3.08\times 10^{-5}$ & [$2.51\times 10^{-5}$, $3.73\times 10^{-5}$] \\
%3-CNOT sequential & $[[66,4,10]]$ & $p_{\mathrm{ebit}}=10^{-2}$ & $1.47\times 10^{-3}$ & 12,314 & 136 & $1.10\times 10^{-2}$ & [$9.27\times 10^{-3}$, $1.31\times 10^{-2}$] \\
%3-CNOT sequential & $[[66,4,10]]$ & $p_{\mathrm{ebit}}=10^{-2}$ & $7.94\times 10^{-3}$ & 122 & 122 & $1.00\times 10^{0}$ & [$9.70\times 10^{-1}$, $1.00\times 10^{0}$] %\\
\hline
3-CNOT sequential & $[[54,4,8]]$ & $p_{\mathrm{ebit}}=10^{-3}$
& $5.00\times 10^{-5}$ & 5,000,000 & 1
& $2.00\times 10^{-7}$
& [$5.06\times 10^{-9}$, $1.11\times 10^{-6}$] \\

3-CNOT sequential & $[[54,4,8]]$ & $p_{\mathrm{ebit}}=10^{-3}$
& $2.71\times 10^{-4}$ & 5,000,000 & 76
& $1.52\times 10^{-5}$
& [$1.20\times 10^{-5}$, $1.90\times 10^{-5}$] \\

3-CNOT sequential & $[[54,4,8]]$ & $p_{\mathrm{ebit}}=10^{-3}$
& $1.47\times 10^{-3}$ & 14,690 & 122
& $8.30\times 10^{-3}$
& [$6.90\times 10^{-3}$, $9.91\times 10^{-3}$] \\

3-CNOT sequential & $[[54,4,8]]$ & $p_{\mathrm{ebit}}=10^{-3}$
& $7.94\times 10^{-3}$ & 110 & 110
& $1.00\times 10^{0}$
& [$9.67\times 10^{-1}$, $1.00\times 10^{0}$] \\
\hline
%\hline
%3-CNOT sequential & $[[66,4,10]]$ & $p_{\mathrm{ebit}}=10^{-3}$ & $5.00\times 10^{-5}$ & 5,000,000 & 0 & $0$ & [$0$, $7.38\times 10^{-7}$] \\
%3-CNOT sequential & $[[66,4,10]]$ & $p_{\mathrm{ebit}}=10^{-3}$ & $2.71\times 10^{-4}$ & 5,000,000 & 68 & $1.36\times 10^{-5}$ & [$1.06\times 10^{-5}$, $1.72\times 10^{-5}$] \\
%3-CNOT sequential & $[[66,4,10]]$ & $p_{\mathrm{ebit}}=10^{-3}$ & $1.47\times 10^{-3}$ & 14,690 & 122 & $8.30\times 10^{-3}$ & [$6.90\times 10^{-3}$, $9.91\times 10^{-3}$] \\
%3-CNOT sequential & $[[66,4,10]]$ & $p_{\mathrm{ebit}}=10^{-3}$ & $7.94\times 10^{-3}$ & 110 & 110 & $1.00\times 10^{0}$ & [$9.67\times 10^{-1}$, $1.00\times 10^{0}$] \\
\end{tabular}%
}
\end{table*}

\begin{table*}[p]
\centering
\caption{Data for the four-node 1-shot GHZ fanout LER-versus-PER graph. The logical error rate is $k/n$, with $k$ observed logical errors in $n$ shots. Confidence limits are exact two-sided 95\% Clopper--Pearson intervals.}
\label{tab:ghz-ler-per}
\scriptsize
\setlength{\tabcolsep}{4pt}
\resizebox{\textwidth}{!}{%
\begin{tabular}{lllr r r r r}
\hline
Method & Code & Parameter & PER & Shots & Errors & LER & 95\% CI for LER \\
\hline
1-shot GHZ & $[[18,4,4]]$ & $p_{\mathrm{GHZ}}=0$ & $5.00\times 10^{-5}$ & 923,238 & 145 & $1.57\times 10^{-4}$ & [$1.33\times 10^{-4}$, $1.85\times 10^{-4}$] \\
1-shot GHZ & $[[18,4,4]]$ & $p_{\mathrm{GHZ}}=0$ & $2.71\times 10^{-4}$ & 34,338 & 117 & $3.41\times 10^{-3}$ & [$2.82\times 10^{-3}$, $4.08\times 10^{-3}$] \\
1-shot GHZ & $[[18,4,4]]$ & $p_{\mathrm{GHZ}}=0$ & $1.47\times 10^{-3}$ & 2,594 & 282 & $1.09\times 10^{-1}$ & [$9.70\times 10^{-2}$, $1.21\times 10^{-1}$] \\
1-shot GHZ & $[[18,4,4]]$ & $p_{\mathrm{GHZ}}=0$ & $7.94\times 10^{-3}$ & 155 & 148 & $9.55\times 10^{-1}$ & [$9.09\times 10^{-1}$, $9.82\times 10^{-1}$] \\
\hline
1-shot GHZ & $[[36,4,6]]$ & $p_{\mathrm{GHZ}}=0$ & $5.00\times 10^{-5}$ & 5,000,000 & 1 & $2.00\times 10^{-7}$ & [$5.06\times 10^{-9}$, $1.11\times 10^{-6}$] \\
1-shot GHZ & $[[36,4,6]]$ & $p_{\mathrm{GHZ}}=0$ & $2.71\times 10^{-4}$ & 2,344,550 & 124 & $5.29\times 10^{-5}$ & [$4.40\times 10^{-5}$, $6.31\times 10^{-5}$] \\
1-shot GHZ & $[[36,4,6]]$ & $p_{\mathrm{GHZ}}=0$ & $1.47\times 10^{-3}$ & 11,554 & 115 & $9.95\times 10^{-3}$ & [$8.22\times 10^{-3}$, $1.19\times 10^{-2}$] \\
1-shot GHZ & $[[36,4,6]]$ & $p_{\mathrm{GHZ}}=0$ & $7.94\times 10^{-3}$ & 118 & 112 & $9.49\times 10^{-1}$ & [$8.93\times 10^{-1}$, $9.81\times 10^{-1}$] \\
\hline
1-shot GHZ & $[[54,4,8]]$ & $p_{\mathrm{GHZ}}=10^{-2}$ & $5.00\times 10^{-5}$ & 4,432,659 & 119 & $2.68\times 10^{-5}$ & [$2.22\times 10^{-5}$, $3.21\times 10^{-5}$] \\
1-shot GHZ & $[[54,4,8]]$ & $p_{\mathrm{GHZ}}=10^{-2}$ & $2.71\times 10^{-4}$ & 955,955 & 115 & $1.20\times 10^{-4}$ & [$9.93\times 10^{-5}$, $1.44\times 10^{-4}$] \\
1-shot GHZ & $[[54,4,8]]$ & $p_{\mathrm{GHZ}}=10^{-2}$ & $1.47\times 10^{-3}$ & 14,642 & 117 & $7.99\times 10^{-3}$ & [$6.61\times 10^{-3}$, $9.57\times 10^{-3}$] \\
1-shot GHZ & $[[54,4,8]]$ & $p_{\mathrm{GHZ}}=10^{-2}$ & $7.94\times 10^{-3}$ & 109 & 107 & $9.82\times 10^{-1}$ & [$9.35\times 10^{-1}$, $9.98\times 10^{-1}$] \\
%\hline
%1-shot GHZ & $[[66,4,10]]$ & $p_{\mathrm{GHZ}}=10^{-2}$ & $5.00\times 10^{-5}$ & 5,000,000 & 77 & $1.54\times 10^{-5}$ & [$1.22\times 10^{-5}$, $1.92\times 10^{-5}$] \\
%1-shot GHZ & $[[66,4,10]]$ & $p_{\mathrm{GHZ}}=10^{-2}$ & $2.71\times 10^{-4}$ & 1,249,523 & 101 & $8.08\times 10^{-5}$ & [$6.58\times 10^{-5}$, $9.82\times 10^{-5}$] \\
%1-shot GHZ & $[[66,4,10]]$ & $p_{\mathrm{GHZ}}=10^{-2}$ & $1.47\times 10^{-3}$ & 14,074 & 113 & $8.03\times 10^{-3}$ & [$6.62\times 10^{-3}$, $9.65\times 10^{-3}$] \\
%1-shot GHZ & $[[66,4,10]]$ & $p_{\mathrm{GHZ}}=10^{-2}$ & $7.94\times 10^{-3}$ & 116 & 116 & $1.00\times 10^{0}$ & [$9.69\times 10^{-1}$, $1.00\times 10^{0}$] \\
\hline
1-shot GHZ & $[[54,4,8]]$ & $p_{\mathrm{GHZ}}=10^{-3}$ & $5.00\times 10^{-5}$ & 5,000,000 & 1 & $2.00\times 10^{-7}$ & [$5.06\times 10^{-9}$, $1.11\times 10^{-6}$] \\
1-shot GHZ & $[[54,4,8]]$ & $p_{\mathrm{GHZ}}=10^{-3}$ & $2.71\times 10^{-4}$ & 4,754,707 & 105 & $2.21\times 10^{-5}$ & [$1.81\times 10^{-5}$, $2.67\times 10^{-5}$] \\
1-shot GHZ & $[[54,4,8]]$ & $p_{\mathrm{GHZ}}=10^{-3}$ & $1.47\times 10^{-3}$ & 19,106 & 122 & $6.39\times 10^{-3}$ & [$5.31\times 10^{-3}$, $7.62\times 10^{-3}$] \\
1-shot GHZ & $[[54,4,8]]$ & $p_{\mathrm{GHZ}}=10^{-3}$ & $7.94\times 10^{-3}$ & 118 & 113 & $9.58\times 10^{-1}$ & [$9.04\times 10^{-1}$, $9.86\times 10^{-1}$] \\
\hline
%1-shot GHZ & $[[66,4,10]]$ & $p_{\mathrm{GHZ}}=10^{-3}$ & $5.00\times 10^{-5}$ & 5,000,000 & 1 & $2.00\times 10^{-7}$ & [$5.06\times 10^{-9}$, $1.11\times 10^{-6}$] \\
%1-shot GHZ & $[[66,4,10]]$ & $p_{\mathrm{GHZ}}=10^{-3}$ & $2.71\times 10^{-4}$ & 5,000,000 & 96 & $1.92\times 10^{-5}$ & [$1.56\times 10^{-5}$, $2.34\times 10^{-5}$] \\
%1-shot GHZ & $[[66,4,10]]$ & $p_{\mathrm{GHZ}}=10^{-3}$ & $1.47\times 10^{-3}$ & 16,826 & 116 & $6.89\times 10^{-3}$ & [$5.70\times 10^{-3}$, $8.26\times 10^{-3}$] \\
%1-shot GHZ & $[[66,4,10]]$ & $p_{\mathrm{GHZ}}=10^{-3}$ & $7.94\times 10^{-3}$ & 119 & 118 & $9.92\times 10^{-1}$ & [$9.54\times 10^{-1}$, $1.00\times 10^{0}$] \\
%\hline
\end{tabular}%
}
\end{table*}

\begin{table*}[t]
\centering
\caption{Logical error-rate data for teleportation using the
$[[36,4,6]]$ code, for the two datasets with $p_{ebit}=10p$ and $p_{ebit}=p$. The LER is the number of logical
errors divided by the number of shots. Confidence intervals are
exact two-sided 95\% Clopper--Pearson intervals.}
\label{tab:36-4-6-teleportation-ler-per}
\scriptsize
\setlength{\tabcolsep}{5pt}
\begin{tabular}{llrrrrr}
\hline
Dataset & PER & Shots & Errors & LER
& \multicolumn{2}{c}{95\% CI for LER} \\
\cline{6-7}
 & & & & & Lower & Upper \\
\hline
$10p$ & $5.00\times10^{-5}$ & 5,000,000 & 2
& $4.00\times10^{-7}$
& $4.84\times10^{-8}$ & $1.44\times10^{-6}$ \\

$10p$ & $2.71\times10^{-4}$ & 1,841,766 & 115
& $6.244\times10^{-5}$
& $5.155\times10^{-5}$ & $7.495\times10^{-5}$ \\

$10p$ & $1.468\times10^{-3}$ & 9,762 & 114
& $1.168\times10^{-2}$
& $9.642\times10^{-3}$ & $1.401\times10^{-2}$ \\

$10p$ & $7.943\times10^{-3}$ & 152 & 129
& $8.487\times10^{-1}$
& $7.817\times10^{-1}$ & $9.016\times10^{-1}$ \\
\hline
$1p$ & $5.00\times10^{-5}$ & 5,000,000 & 0
& $< 2.00 \times 10^{-7}$$^{*}$
& $0$ & $7.378\times10^{-7}$ \\

$1p$ & $2.71\times10^{-4}$ & 4,479,078 & 112
& $2.501\times10^{-5}$
& $2.059\times10^{-5}$ & $3.009\times10^{-5}$ \\

$1p$ & $1.468\times10^{-3}$ & 15,906 & 118
& $7.419\times10^{-3}$
& $6.144\times10^{-3}$ & $8.878\times10^{-3}$ \\

$1p$ & $7.943\times10^{-3}$ & 170 & 123
& $7.235\times10^{-1}$
& $6.498\times10^{-1}$ & $7.893\times10^{-1}$ \\
\hline
\end{tabular}\\
* This is due to zero errors in 5,000,000 shots though with more shots, errors could emerge.
\end{table*}

\end{document}